\documentclass[doublecol]{epl2} 

\usepackage{graphicx}
\usepackage{graphics}
\usepackage{epsfig}
\usepackage{color}
\usepackage{amssymb}
\usepackage{amsmath}
\usepackage{bm}
\usepackage{wasysym}
\usepackage{pifont}

\newcommand{\be}{\begin{equation}}
\newcommand{\ee}{\end{equation}}
\newcommand{\ba}{\begin{eqnarray}}
\newcommand{\ea}{\end{eqnarray}}

\title{Critical scaling and heterogeneous superdiffusion across the jamming/rigidity transition of a granular glass}
\shorttitle{Critical jamming/rigidity transition} 

\author{F. Lechenault\inst{1} \and O. Dauchot\inst{1} \and G. Biroli\inst{2} \and J.~P. Bouchaud\inst{3} }
\shortauthor{F. Lechenault \etal}

\institute{                    
  \inst{1} CEA Saclay/SPEC, URA2464, L'Orme des Merisiers, 91 191 Gif-sur-Yvette, France\\
  \inst{2} CEA Saclay/SPhT, UMR2306, L'Orme des Merisiers, 91 191 Gif-sur-Yvette, France\\
  \inst{3} Science \& Finance, Capital Fund Management, 6-8 Bd Haussmann, 75009 Paris, France
}

\abstract{
The dynamical properties of a dense horizontally vibrated bidisperse granular monolayer are experimentally investigated. The quench protocol produces states with a frozen structure of the assembly, but the remaining degrees of freedom associated with contact dynamics control the appearance of macroscopic rigidity. We provide decisive experimental evidence that this transition is a critical phenomenon, with increasingly collective and heterogeneous rearrangements occurring at length scales much smaller than the grains' diameter, presumably reflecting the contact force network fluctuations. Dynamical correlation time and length scales soar on both sides of the transition, as the volume fraction varies over a remarkably tiny range ($\delta \phi/\phi \sim 10^{-3}$). We characterize the motion of individual grains, which becomes super-diffusive at the jamming transition $\phi_J$, signaling long-ranged temporal correlations. Correspondingly, the system exhibits long-ranged four-point dynamical correlations in space that obey critical scaling at the transition density.
}

\begin{document}

\maketitle

\section{Introduction}
As the volume fraction of hard grains is increased beyond a certain point, the system jams and is able to support mechanical stresses. It has been argued that this rigidity transition is akin to phase transitions in thermal systems, and that the jamming density is a genuine critical point~\cite{Jamming}. Despite recent efforts~\cite{Jamming,Thorpe,Wyart,VanHecke,D'Anna,Pouliquen,Kabla,Marty,Reis,AbateDurian,weeks}, the mechanisms underlying the jamming transition remain an open problem. ``Jamming'' appears to be associated with two different notions, not always well distinguished in the literature. On the one hand, the {\it glass/jamming} transition reflects the freezing of {\it structural} degrees of freedom associated to topological neighborhoods and the divergence of the structural relaxation time $\tau_{\alpha}$, as observed in glass-forming liquids, colloidal suspensions and granular assemblies. The common mechanism for the divergence of $\tau_\alpha$ appears to be increasingly collective rearrangements~\cite{Dauchot,ScienceBerthieretal,Abate,luca}, supporting the idea that the transition is a critical phenomenon -- albeit of a new kind. On the other hand, the {\it rigidity/jamming} transition, is the appearance of mechanical rigidity~\cite{Jamming,Wyart,VanHecke,Behringer} for which there are only a few direct experimental evidence of the mechanisms at play. Contrary to the glass/jamming transiton, it does not require the system to be at equilibrium: an assemblies of grains will eventually becomes rigid at high enough density regardless of the rate of squeezing (an extremely small rate corresponds to the ``equilibrium route''). This rigidity/jamming transition may in principle be different from the glass/jamming transition as indeed found in mean-field glassy models \cite{jorge,semerjian}, in a previous experiment \cite{Swinney} and below.
 
\begin{figure}[t] 
\center
\includegraphics[width=0.4\textwidth,height=3cm]{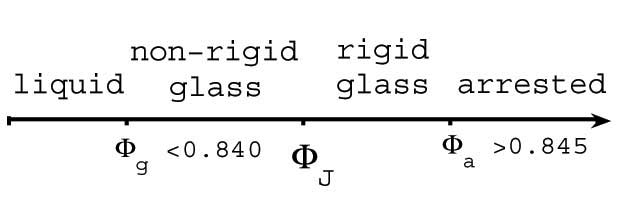}
\vspace{-0.2cm}
\caption{The packing fraction axis and the rigidity transition with respect to the glass transition and the fully arrested state. Previous work on dynamical heterogeneities in granular packings concerned the liquid region, approaching the glass transition~\cite{Marty,AbateDurian,Dauchot,Abate}, while the present work focuses on $\phi > \phi_g$, the glass transition packing fraction, which for our experimental timescale is less than $0.84$.}
\label{fig:schema}
\vspace{-0.0cm}
\end{figure}

In this letter, we investigate the rigidity/jamming transition in an assembly of horizontally vibrated bi-disperse hard disks. In contrast with previous studies~\cite{Marty,AbateDurian,Dauchot,Abate}, we use a quench protocol that produces very dense states, far beyond the glass density $\phi_g$ relative to our experiment, i.e. such that the structural relaxation time $\tau_\alpha$ is much larger than the experimental time scale. However, there is a density range $\phi_g <\phi < \phi_a$ where the strong vibration can still induce micro-rearrangements through collective contact slips that lead to a partial exploration of the portion of phase space, restricted to a particular frozen structure. For $\phi > \phi_a$, on the other hand, the system is completely arrested (see Fig.~\ref{fig:schema}). We study how these structurally frozen, but still evolving states acquire or lose their rigidity as the packing fraction crosses the value $\phi_J \approx 0.842$. We (i) find that the collective contact rearrangements build, around $\phi_J$, non trivial diffusion properties at length scales much smaller than the grain diameter, (ii) demonstrate that the associated microscopic motion occurs through persistent, but disordered large-scale currents and (iii) provide the first direct experimental evidence of a simultaneous divergence of dynamical correlation length and time scales at the rigidity/jamming transition $\phi_J$, distinct from those of the glass/jamming transition, and presumably associated with the underlying heterogeneous dynamics of the force network.  

\section{Experimental set-up and protocol}
\begin{figure}[t] 
\center
\includegraphics[width=0.4\textwidth,height=3cm]{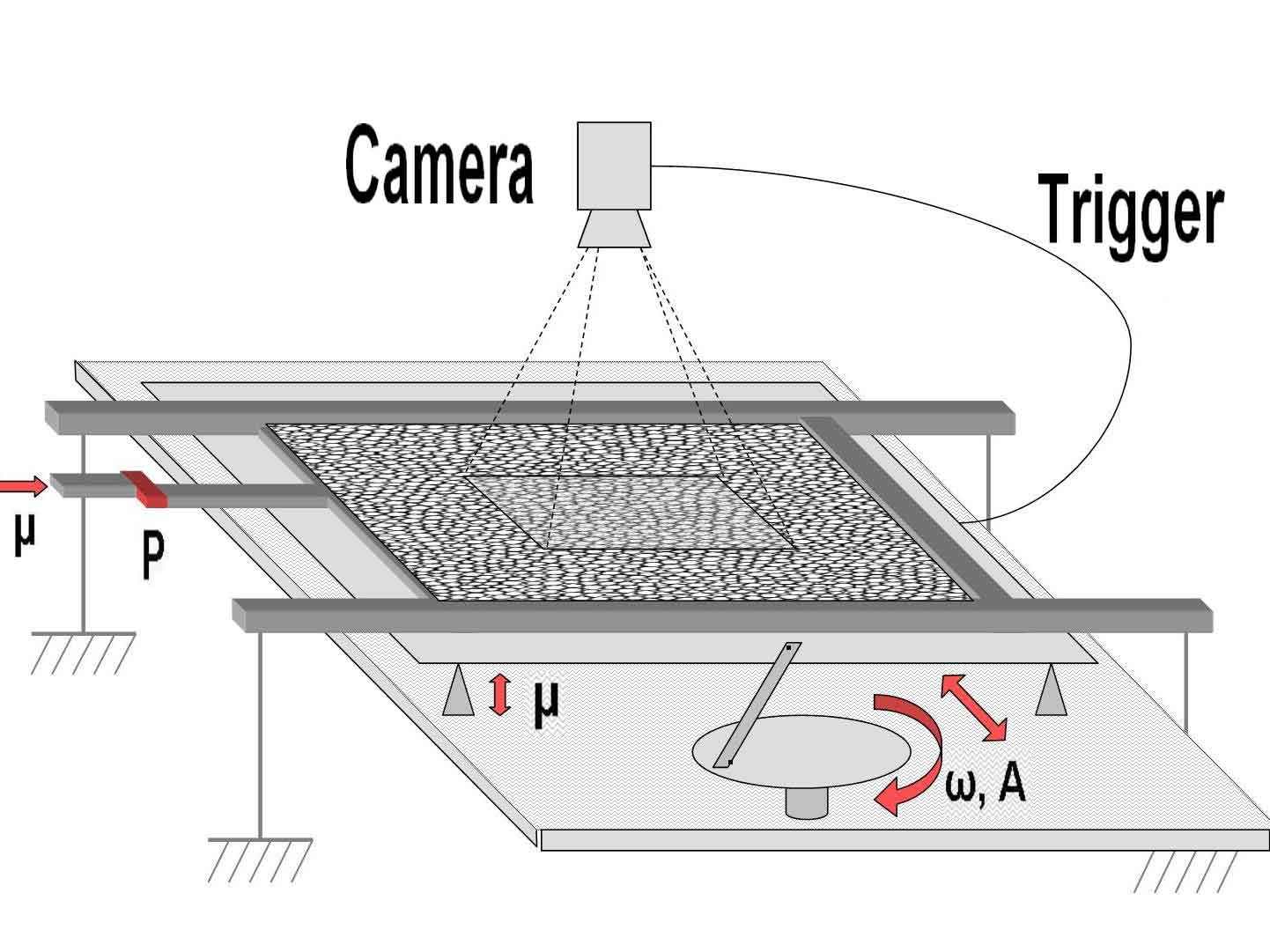}
\vspace{-0.2cm}
\caption{Experimental system: one monolayer of disks is vibrated horizontally, while controlling the volume fraction and tracking the grains}
\label{fig:setup}
\vspace{-0.0cm}
\end{figure}

Our experimental system is shown in fig.~\ref{fig:setup}. A 1:1 bidisperse monolayer of 8500 brass cylinders of diameters $d_{small} = 4\pm0.01 mm$ and $d_{big} = 5\pm0.01 mm$ laid out on a horizontal glass plate mounted on three micro-controlled stands for accurate level control. Power injection is provided by a $1.5$ kW motor with $4.77$ N.m nominal torque, bond to an eccentric rotor and rod system that oscillates the plate horizontally in one direction at a frequency of 10 Hz and with a peak-to-peak amplitude of 10 mm. The grains are confined within a fixed rectangular metal frame while the bottom plate moves, so that energy is homogeneously injected in the bulk of the grain assembly through friction. The cell has width L $\approx 100$ $d_{small}$, and its length can be adjusted by a lateral mobile wall controlled by a $\mu m$ accuracy motorized translation stage, which allows us to vary the packing fraction of the grains assembly -- the control parameter for our experiment -- by tiny amounts ($\delta \phi/\phi \sim 5 \, 10^{-4})$ within an accuracy of $10^{-4}$. The pressure exerted on this wall is measured by a force sensor inserted between the wall and the stage. The stroboscopic motion of a set of 1500 grains in the center of the sample is tracked with an accuracy of $2.10^{-3} d$  by a CCD camera triggered through TTL electronics by a reflector mounted on the rotor and probed by an infrared sensor. Lengths are measured in $d_{small}$ units and time in cycle units.

\begin{figure}[t] 
	\begin{minipage}[c]{.49\linewidth}
	\includegraphics[width=1\textwidth]{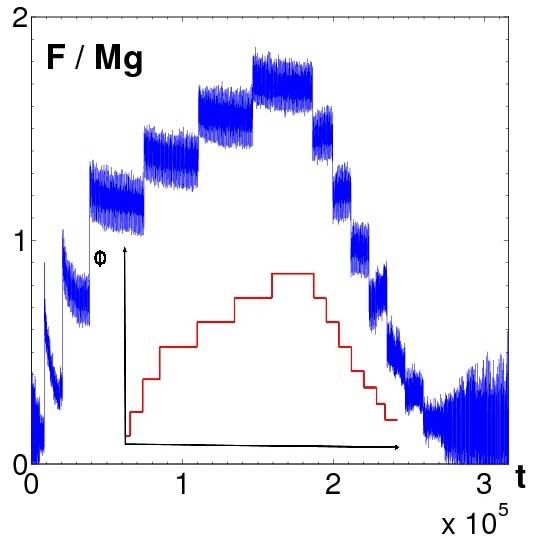}
	\end{minipage}
	\begin{minipage}[c]{.49\linewidth}
	\vspace{-0.2cm}
	\includegraphics[width=1\textwidth]{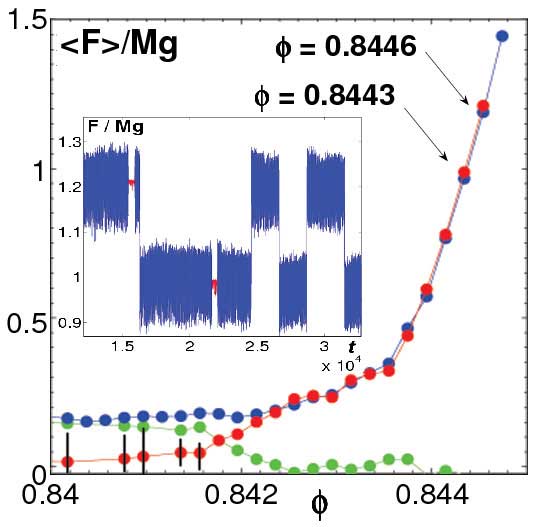}
	\end{minipage}
\vspace{-0.2cm}
\caption{\textbf{Left}: Normal stress on the side of the cell normalised by the total grains weight, as a function of time during a typical run. \textbf{Inset}: Packing fraction protocol. \textbf{Right}: Mean (blue), mean static (red) and mean kinetic (green) constraints as a function of packing fraction. Vertical segments crossing the red circles indicate the range of fluctuations. \textbf{Inset}: Density cycles between the two values of the packing fraction pointed out by the arrows indicate a reversible and stationary pressure signal. The data drawn in red corresponds to periods of time during which we stop the vibration and measure the residual static constraint.}
\label{fig:SDP}
\vspace{-0.0cm}
\end{figure}

In order to reach a steady state analogous to those reported in~\cite{Nagel}, i.e. in which macroscopic observables are in one to one correspondence with the control parameter in a reversible and history independent manner, and provide reproducible measurements, we first perform a series of compression steps to reach a highly jammed configuration with  $\phi=0.8457$, before we stepwise decompress the system. The pressure signal is acquired as a function of time (both with and without vibration) and is shown in Fig.\ref{fig:SDP}. In the compression phase, the pressure relaxes after each step but never reaches a stationary state. In the decompression phase however, stationary (no aging was observed) and reversible  (see compression cycles on Fig.\ref{fig:SDP}) pressure signals are obtained. However, the structural relaxation time $\tau_{\alpha}$ is much longer than our experimental time. Actually, the neighbors of a given particles typically  do not change on our experimental timescales, indicating that we are focusing on a regime where $\phi > \phi_g$. Although the system is not fully equilibrated, it appears to be in partial equilibrium {\it within} the basin of configurations corresponding to a given structural arrangement of the grains. The mean pressure, its residual static part (when vibration is switched off) and the kinetic part defined as their difference are shown as a function of the packing fraction in Fig.\ref{fig:SDP}, in the decompression phase. At high packing fractions, static internal stresses are important, and completely dominate the mean pressure. The grain assembly, in this regime, is rigidly locked to the frame. As the packing fraction is decreased, a kinetic contribution appears as the grains unlock from the side walls and start making a distinctive noise. We identify these phenomena with the rigidity/jamming transition, and thereby obtain $\phi_J \in [0.8417, 0.8422]$. Whereas the precise value of $\phi_J$ might depend on the details of the protocol, the critical features observed at the transition are expected to be universal. Nevertheless, we note that the value $\phi_J \approx 0.842$ is remarkably close to $\phi_J=0.8422$ reported  in~\cite{Behringer}, using different grains and a different experimental protocol. At lower packing fractions, the pressure is purely of kinetic origin. The grains slide on the sides and moves along with the plate, and the static part is zero.

\section{Experimental results}
We now analyze the evolution of the dynamics of the grains, namely the mean square displacement, the self-intermediate scattering function, and the dynamical correlations, as a function of the packing fraction {\it across} the rigidity/jamming transition. 

\begin{figure}[t]
	\begin{minipage}[c]{.49\linewidth} \centering
	\vspace{-0.15cm}
	\includegraphics[width=1\textwidth,height=4.2cm]{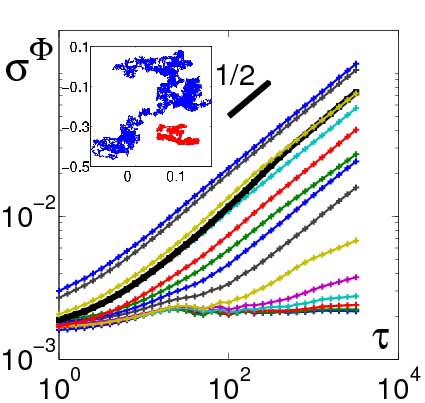}
	\end{minipage}
	\begin{minipage}[c]{.49\linewidth} \centering 
	\includegraphics[width=0.97\textwidth,height=4.1cm]{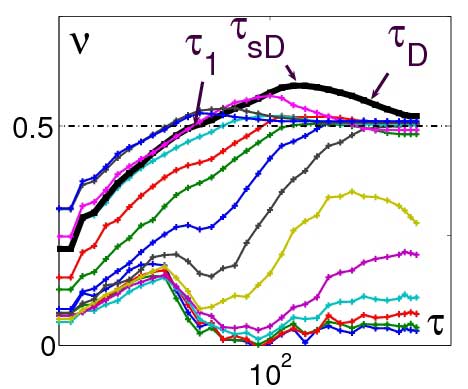}
	\end{minipage}
\caption{\textbf{Left}: $\sigma_{\phi}$ as a function of $\tau$ for $\phi=0.8402$, $0.8407$, $0.8413$, $0.8417$, $0.8422$, $0.8426$, $0.8430$, $0.8435$, $0.8440$, $0.8443$, $0.8446$, $0.8450$, $0.8452$, $0.8454$. Increasing values of $\phi$ tends to reduce $\sigma_{\phi}(\tau)$ (except around $\phi_J$). The bold line corresponds to the packing fraction $0.8417$ closest to $\phi_{J}$ -- note that it {\it crosses} the data corresponding to $\phi <\phi_J$. \textbf{Inset}: Typical trajectories of a grain during a 20 minute acquisition at packing fractions smaller (blue) and larger (red) than $\phi_{J}$. \textbf{Right} Local slope $\nu = \partial \log \sigma_{\phi}(\tau)/\partial \log(\tau) $ vs. $\tau$ for the same packing fractions.}
\label{fig:diff}
\vspace{-0.0cm}
\end{figure}

Ratting particles, identified as grains performing instantaneous jumps much larger than the average motion, exhibit a strongly anisotropic dynamics, but appear to be a rather marginal population - ranging from 1 to 5 percent of the grains respectively in the highest and lowest density states. All dynamical quantities are hence computed for non-rattling particles. As a result, and even though the macroscopic vibration is unidirectional, motion is found to be isotropic -- a feature already observed in~\cite{Marty}. We compute the mean square displacement for a density $\phi$ and lag~$\tau$:
\be 
\sigma^2_{\phi}(\tau)\equiv \left\langle \frac{1}{N}\sum_{i} 
\left\|\Delta\vec r_{i}\right\|^2(t,\tau)\right\rangle_{t} 
\ee
\noindent where $N$ is the number of imaged grains and $\left\|\Delta\vec r_{i}\right\|^2(t,\tau)$ is the square displacement of grain $i$ between times $t$ and $t+\tau$ in the grains' geometric center of mass frame. The time average $\langle \cdot \rangle_t$ is performed over $10,000$ cycles taken $1000$ cycles after each step in $\Phi$. The evolution of $\sigma^2_{\phi}(\tau)$ with the packing fraction is presented in Fig.\ref{fig:diff}. Note the very small values of $\sigma_{\phi}(\tau)$ at all timescales, which is a further indication that the packing indeed remains in a given structural arrangement for the range of densities and time scales we are considering here. Although very small, $\sigma_{\phi}(\tau)$ is well above experimental error bars
and is {\it not} the result of noise, as the following analysis further demontrates.
At low packing fractions $\phi < \phi_J$, and at small $\tau$ the mean square displacement displays a sub-diffusive behavior. At longer time, a diffusive regime is observed, at least up to our experimental time scale. As the packing fraction is increased, the typical lag at which this cross-over occurs becomes larger and, at first sight, does not seem to exhibit any special feature for $\phi\simeq\phi_J$ (corresponding to the bold line in Fig.\ref{fig:diff}). Above $\phi_J$, an intermediate plateau appears before diffusion resumes. However, a closer inspection of $\sigma^2_{\phi}(\tau)$ reveals an intriguing behavior, that appears more clearly on the local logarithmic slope $\nu={\partial \log \sigma_{\phi}(\tau)}/{\partial \log(\tau)}$ shown in Fig.\ref{fig:diff}. When $\nu=\frac{1}{2}$, the motion is diffusive, whereas at small times, $\nu < \frac{1}{2}$ indicates sub-diffusive behavior. At intermediate packing fractions, instead of reaching $\frac{1}{2}$ from below, $\nu$ overshoots and reaches values $>\frac{1}{2}$ before reverting to $\frac{1}{2}$ from above at long times. Physically, this means that after the sub-diffusive regime, the motion becomes {\it super-diffusive} at intermediate times before eventually decorrelating and entering the diffusive regime. At higher packing fractions, this intermediate superdiffusion disappears: one only observes a crossover between a plateau regime at early times and diffusion at ``long'' times (but still small compared to $\tau_\alpha$: as noted above, we only explore the regime where $\sigma_\phi \ll d$).

\begin{figure}[t]
\centering
\includegraphics[width=0.4\textwidth,height=4.5cm]{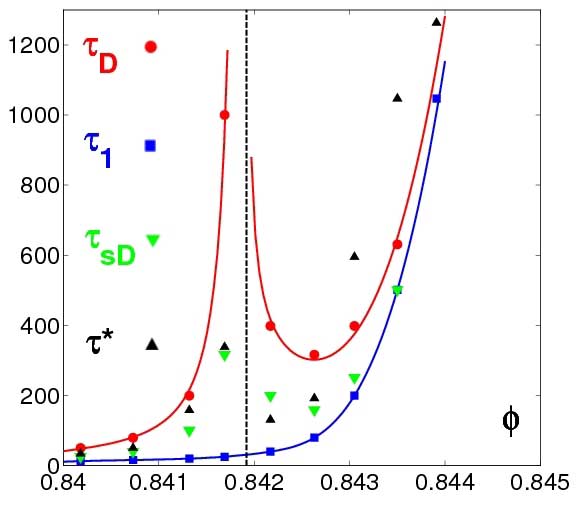}
\caption{Timescales $\tau_1, \tau_{sD}, \tau_{D}$ and $\tau^*$  as a function of the packing fraction. The lines are guides to the eyes obtained by best power-law fits. The dashed line indicates the packing fraction $\phi_J$ corresponding to the jamming transition}
\label{fig:times}
\vspace{-0.0cm} 
\end{figure}

In order to characterize these different regimes, we define three characteristic times: $\tau_1(\phi)$ as the lag at which $\nu(\tau)$ first reaches $1/2$, corresponding to the start of the super-diffusive regime, $\tau_{sD}(\phi)$ when $\nu(\tau)$ reaches a maximum $\nu^{*}(\phi)$ (peak of super-diffusive regime), and $\tau_{D}(\phi)$ where $\nu(\tau)$ has an inflection point, beyond which the system approaches the diffusive regime. These characteristic time scales are plotted as a function of the packing fraction in Fig.\ref{fig:times}. Whereas $\tau_1$ does not exhibit any special features across $\phi_J$, both $\tau_{sD}$ and $\tau_{D}$ are strongly peaked at $\phi_J$. Since superdiffusion is tantamount to long-time correlations in the motion of particles, this result shows that dynamical correlations are maximal at $\phi_J$, reinforcing the interpretation of $\phi_J$ as a critical packing fraction.

\begin{figure}[t]
	\begin{minipage}[c]{.49\linewidth} \centering
	\includegraphics[width=1\textwidth,height=4.1cm]{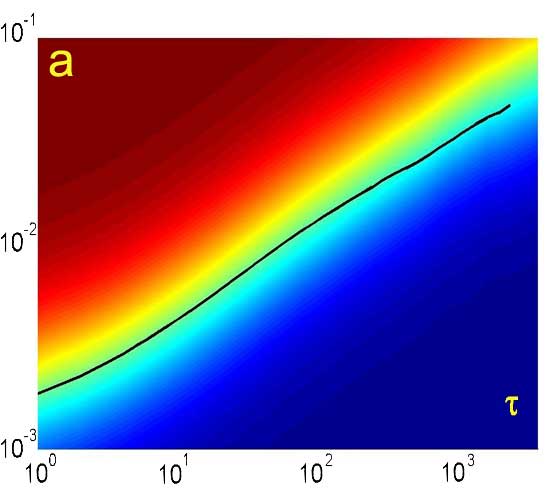}
	\end{minipage}
	\begin{minipage}[c]{.49\linewidth} \centering 
	\includegraphics[width=1\textwidth,height=4.1cm]{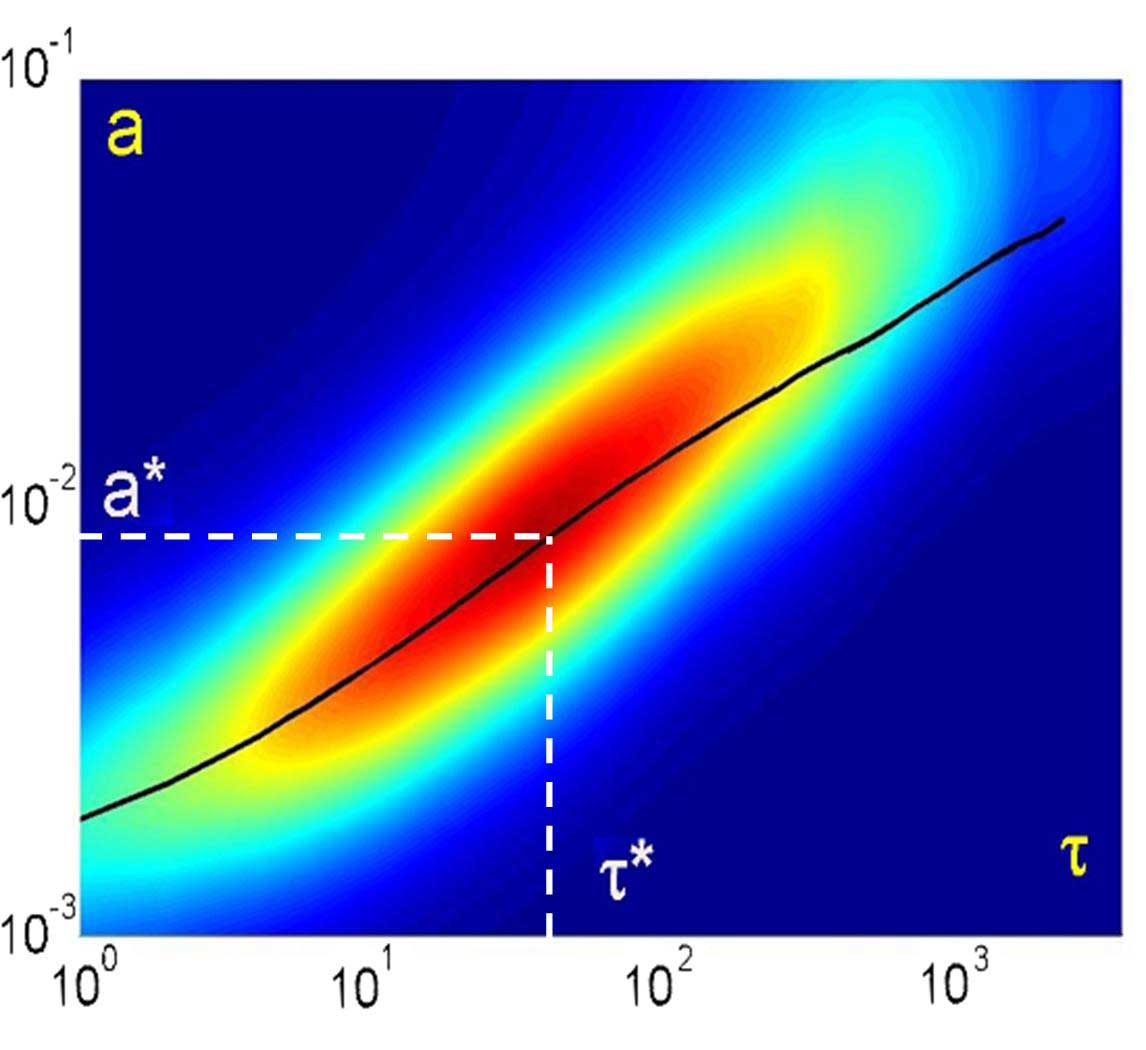}
	\end{minipage}
\caption{\textbf{Left}: $Q_a(\tau)$ and \textbf{Right}: $\chi_{4,a}(\tau)$ as a function of probing length $a$ and delay time $\tau$ for $\phi=0.8413<\phi_{J}$ in a log-log color plot. The bold line represents $0.613~\sigma_{\phi}(\tau)$.}
\label{fig:QXiatau}
\vspace{-0.0cm}
\end{figure}

As a decisive signature of the critical nature of $\phi_J$, we now show that these long-time correlations are accompanied by the growth of spatial correlations of the dynamics. We follow previous works~\cite{FranzParisiBerthierHarrowell}, where the four-point correlation $G_4(\vec r,\tau)$, and its integral over space $\chi_4(\tau)$ were proposed to measure the fluctuations of the temporal relaxation (see also~\cite{Dauchot,Abate}). We characterize the local dynamics through the correlation function $Q_a(\tau)\equiv\left\langle Q^t_a(\vec x,\tau) \right\rangle_{t,x}$:
\begin{equation}
Q^t_a(\vec x,\tau) \equiv \frac{1}{N}\sum_{i} \delta(\vec r_i(t) - \vec x) \, \exp{\left(-\frac{\left\|\Delta\vec r_{i}\right\|^2}{2a^2}\right)} 
\end{equation}
where $a$ is the length scale over which we probe the motion. $Q_a(\tau)$ is akin to the self-intermediate scattering function in glass-forming liquids. An example of $Q_a(\tau)$ for a specific value of $\phi$ is shown in the left panel of Fig.\ref{fig:QXiatau} as a function of probing length $a$ and lag $\tau$. The bold line represents $\tilde a\equiv\lambda \sigma_{\phi}(\tau)$, where $\lambda=0.613$ is a density-independent constant. It is found to define the crest where $Q_a(\tau)$ decreases from one -- when no significant motion has occurred -- to zero -- when all grains have made a displacement larger then $a$. We found $Q_a(\tau)$ to obey the scaling relation  $Q_a(\tau) = \tilde Q(\sigma_{\phi}(\tau)/a)$ at all $\phi$, showing that $\sigma_{\phi}(\tau)$ rescales the whole distribution of displacements at lag $\tau$, hence being the only microscopic length scale relevant to {\it individual} motion at this time scale. The four-point dynamical correlation is defined as: 
\be G_{4,a}(\vec r,\tau) \equiv \left\langle Q^t_a(\vec x+\vec r,\tau) Q^t_a(\vec x,\tau)\right\rangle_{t,x} -  \left\langle Q^t_a(\vec x,\tau)\right\rangle_{t,x}^2. 
\ee

\begin{figure*}[t]
\includegraphics[width=0.33\textwidth,height=5cm]{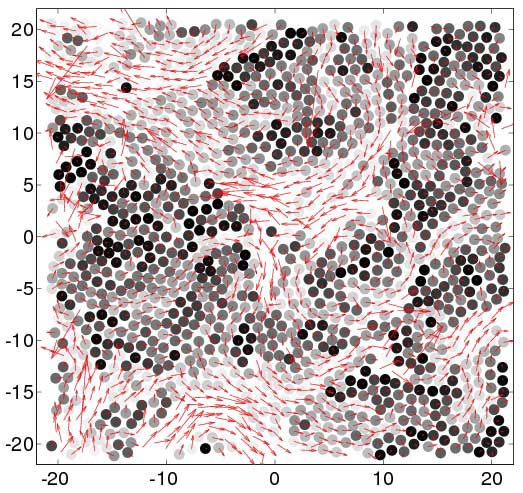}
\includegraphics[width=0.33\textwidth,height=5cm]{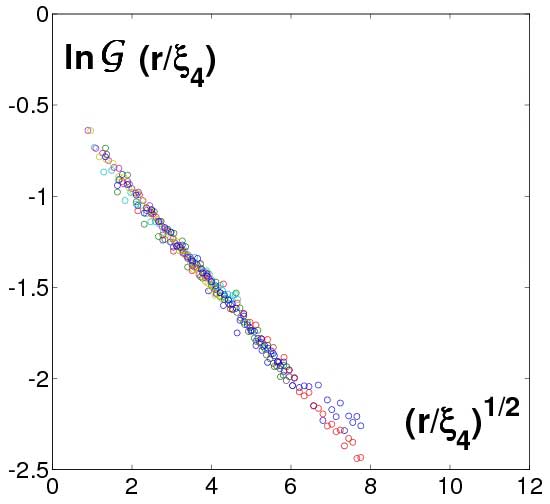}
\includegraphics[width=0.33\textwidth,height=5cm]{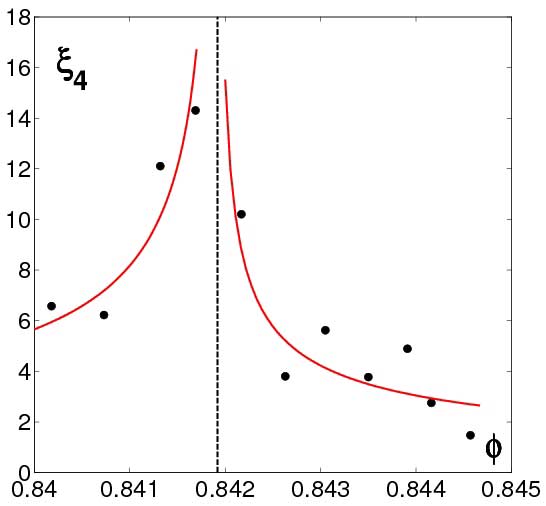}	 
\caption{\textbf{Left}: Typical snapshot of the $Q_a^t(\vec{x},\tau^*)$ at $\phi_J$ and the corresponding displacement field (magnified 5 times). \textbf{Center}: Rescaling of $\ln [G_{4,a^*}(\vec r,\tau^*)/G_{4,a^*} (0,\tau^*)]$ as a function of $\sqrt{r/\xi_4}$ for 8 densities around $\phi_J$. \textbf{Right}: $\xi_{4}$ as a function of the packing fraction. The lines are indicative power-laws $ \sim A_\pm/|\phi - \phi_J|^{1/2}$ consistent with our data, but not an experimental determination of critical exponents.}
\label{fig:chi}
\vspace{-0.0cm} 
\end{figure*}
 
\noindent
The susceptibility $\chi_{4,a}(\tau)=\int d\vec r G_{4,a}(\vec r,\tau)$ gives access to a typical number of grains that move in a correlated way when the system evolves over a given time scale $\tau$. An example of $\chi_{4,a} (\tau)$ for the same specific value of $\phi$ is shown in the right panel of Fig.\ref{fig:QXiatau} together with the same bold line as above. This time it defines the crest where $\chi_{4,a}(\tau)$ is locally maximum: for any given $a$, $\chi_{4,a}(\tau)$ is very small both when $\tau \to 0$, when no significant motion has occurred, and at large $\tau$ corresponding to homogeneous normal diffusion. Hence $\chi_{4,a}$ partly inherits the strong scaling property found above for $Q_{a}$ and reads $\chi_{4,a} (\tau) = h_{\phi}\left(\tau\right)\tilde\chi\left(\sigma_{\phi}(\tau)/a\right)$ where the amplitude $h_{\phi}$ encodes the relevant collective information. The global maximum  $\chi_{4}^{*}(\phi)$ is reached for a density dependent lag $\tau^*(\phi)$ and a probing length $a^{*}(\phi)=\lambda \sigma_{\phi}\left(\tau^*\right)$ for which the dynamics is maximally heterogeneous. Thus $a^*$ it is not fixed from the beginning, as often the case in the recent literature on $\chi_4$, but is rather imposed by the dynamics itself. We found that $\tau^{*}$ is always very close to $\tau_{sD}$ (at least in the vicinity of $\phi_J$), thus indicating that dynamical correlations are maximal when the dynamics is maximally super-diffusive. A typical snapshot of the $Q_a^t(\vec{x},\tau^*)$ at $\phi_J$ and the corresponding displacement field are shown in the left panel of Fig.\ref{fig:chi}; it reveals that the dynamical heterogeneity is in the present case organized in channel currents meandering between blobs of blocked particles. These events occur on time scales such that $\tau^*\approx \tau_{sD}$. Hence the observed currents are persistent in time, and they build up the anomalous super-diffusive motion observed in our system -- and also reported in soft disks simulations~\cite{LiuGoree} -- even though the amplitude of the corresponding displacements is in the present case a very small fraction of a grains diameter. More strikingly, we have found that the four-point correlation function $G_{4,a^*} (\vec r,\tau^*)$ obeys critical scaling in the vicinity of $\phi_J$ (see central panel of Fig.\ref{fig:chi}): 
\be
G_{4,a^*}(\vec r,\tau^*) \propto \frac{G_{4,a^*}(0,\tau^*)}{r^{\alpha}} {\cal G}\left(\frac{r}{\xi_4}\right)
\ee 
with ${\cal G}(u) = \exp(-u^{-\eta})$, $\eta \approx - \frac{1}{2}$ and where $\xi_4(\phi)$, plotted in the right panel of Fig.\ref{fig:chi}, is the length scale over which dynamical correlations develop. The value of exponent $\alpha$ is found to be smaller than $0.15$ but cannot be determined accurately.  This scaling form, together with the strong increase of both $\xi_4(\phi)$ and $\tau^{*}(\phi)$ over a minute range of $\phi$, is our strongest evidence that the jamming fraction $\phi_J$ is indeed a critical point, where a static pressure appears {\it and} long-range dynamical correlations develop. 

\section{Discussion}
As emphasized above, the typical displacements are very small compared to the grain size: $\sigma_\phi(\tau_D) \ll d$. Rather, these length scales compare to inter-particle spacing and typical slip events. The topological neighborhoods do not evolve: when $\phi > \phi_g$ all experimental time scales are shorter than the characteristic structural relaxation time $\tau_{\alpha}$. The exploration of phase space hence probes different {\it contact configurations} rather than different {\it structural arrangements}~\cite{leshouches}. 

The length scale, or wave-vector, at which the dynamics is probed is varied continuously over more than two decades (see \cite{garrahan} for a similar analysis in a model of glass-forming liquid). This analysis reveals that the global maximum $\chi_{4}^{*}(\phi)$ of the four point susceptibility is reached for length and time scales $a^{*}(\phi)$ and  $\tau^*(\phi)$ such that $a^{*}(\phi)\approx\sigma_{\phi}(\tau^*)$. This implies two important consequences. First, the collective dynamics occur on length and time scales selected by the dynamics itself and are not arbitrarily chosen. Second, it was shown for glass forming liquids that the dynamical susceptibility $\chi_{4,a}$ obeys the following general inequality: $\chi_{4,a} \ge ({\partial Q_a}/{\partial \phi})^2 \langle \phi ^2 \rangle_c$ \cite{ScienceBerthieretal}. Plugging in the scaling behaviors $Q_a(\tau;\phi) = \tilde Q\left(\sigma_{\phi}\left(\tau\right)/a\right)$ and $\chi_{4,a} (\tau) = h_{\phi}\left(\tau\right)\tilde\chi\left(\sigma_{\phi}(\tau)/a\right)$, one finds~\cite{EPLBound} that the temporal variations of $\chi_{4,a}$, in particular the presence of a global maximum, are in fact completely encoded in the anomalous behavior of $\sigma_{\phi}(\tau)$, proving the strong connection between anomalous diffusion and dynamical correlations.

Altogether, the jamming/rigidity is found to be a critical phenomenon characterized by the emergence of giant dynamical fluctuations at the contact scale, embodied by extended and collective slip events. We speculate that these dynamical fluctuations mirror the underlying heterogeneous force network and its dynamics, since the contact network is more likely to slip in the regions of concentrated constraint gradients, i.e. along force chains. The mechanism by which diffusion sets in is therefore through collective contact slips allowed by force fluctuations. The decorrelation time $\tau_D$ corresponds in this picture to the time it takes for the dynamics to achieve a complete renewal of the force network. Thanks to these micro-rearrangements, the effective compression modulus of the system under vibration turns out to be orders of magnitude smaller than that of individual grains, which can be considered as infinitely rigid. We believe that friction plays an important role here, by allowing the system to sustain an external pressure at densities smaller than \textbf{$\phi_a$}. It would be interesting to elaborate on Edwards' ideas~\cite{EdwardsBertin} and understand quantitatively the shape of the pressure vs. density observed in Fig.\ref{fig:SDP} which appears to scale as $P \sim (\phi-\phi_J)$ close to $\phi_J$, as also reported in~\cite{Behringer}.

Hence, at variance with what is often assumed, the rigidity/jamming fraction $\phi_J$ is not the point $\phi_g$ at which the structure freezes within the experimental time scales. In fact, studies of glassy systems within a sort of mean field approximation suggest that the ideal glass (or glass/jamming) transition is distinct from the rigidity/jamming transition \cite{jorge,semerjian,zamponi} which is not either the point $\phi_a$ at which dynamics is totally arrested, as found by us and also reported in ~\cite{Swinney}: an {\it homogeneous} diffusive regime is found after sufficiently long times both below and above $\phi_J$ (at least up to the longest available times). Only after further compression will the characteristic time, $\tau_1$, diverge, leading to a complete arrest of the system. In our system, we found $\phi_a > 0.845$, a lower bound for the random close packing fraction. When $\phi_J < \phi < \phi_a$, even though the system is mechanically rigid, the vibration enables it to explore different micro-configurations. This exploration eventually generates uncorrelated particles displacements on longer time scales, hence diffusive motion at a very small scale. What will happen at time scales much longer than the experimental ones?  Depending on the volume fraction, either the dynamics starts feeling the frozen structure and subdiffusion sets in again or the structure unfreezes and diffusion is maintained for ever. Predicting such very long term evolution amounts to address the question of the existence of an ideal Kauzmann-like glass/jamming transition and its relation to the rigidity transition. Such a conceptual issue is obviously far beyond the scope of the present study.

Following recent work~\cite{Wyart}, it is tempting to conjecture that the correlated currents observed here are related to the extended {\it soft modes} that appear when the system loses or acquires rigidity at $\phi_J$. Under the action of a mechanical drive the system should fail along these soft modes. This picture is also compatible with the Mode-Coupling Theory (and some Kinetically Constrained Models), where deep metastable states become marginally stable at the transition ~\cite{Grigera,SellittoBiroliToninelli}, with a diverging dynamical correlation length ~\cite{BiroliBouchaud,Schwarz}. Since in our case $\tau_\alpha \gg \tau^*$, dynamics is confined within a {\it unique} structural arrangement and hence activated events are absent, as assumed in MCT. More work is needed to confirm (or disprove) these ideas and rationalize the value of the critical exponents $\alpha$, $\eta$ and $\nu^{*}(\phi_J)$ associated to the spatio-temporal correlations of these structures, as well as the critical exponents governing the divergence of $\tau^{*}$ and $\xi_{4}^{*}$ at $\phi_J$. The link with the critical properties at the glass/jamming transition (as studied in \cite{Abate} for looser systems $\phi \sim \phi_g$) also needs to be clarified.

In this work, we have provided convincing evidence that the rigidity/jamming transition is a critical point, where long range correlations develop both in time and space. The jamming density, $\phi_J$, where mechanical rigidity sets in, is found to be distinct from the density $\phi_a$ where the dynamics is fully arrested; we suspect that both friction and drive are responsible for this intermediate regime. The loss of rigidity results from the emergence of large scale heterogeneous superdiffusive currents. Remarkably the individual displacements themselves are much smaller than the grain size pointing out the role of the force network rearrangments. We expect that our findings may be significant to understand the dynamics and rheology of dense granular media, colloids and other soft glassy materials, where correlated clusters of mobile particles have been reported on several occasions, in particular in connection with soft modes. It may also well be that the seemingly ubiquitous `compressed exponential' relaxation in soft glassy materials ~\cite{luca} is related to the superdiffusive motion unveiled here~\cite{pitard}.

We would like to thank L. Berthier, M. Bonetti, M. Cates, L. Cipelletti, D. Reichman, M. van Hecke, and M.Wyart for helpful discussions and comments. We also thank V. Padilla and C. Gasquet for technical assistance.


\begin{thebibliography}{0}

\bibitem{Jamming} C. S. O'Hern, S.A. Langer, A. J. Liu and S. R. Nagel, {\it Phys. Rev. Lett.} {\bf 88} 075507 (2002).
\bibitem{Thorpe} F.M. Thorpe, P.M. Duxbury, {\it Rigidity Theory and Applications}, Kluwer Academic/Plenum Publishers, (1999).
\bibitem{Wyart} M. Wyart, S. R. Nagel and T. A. Witten, Europhys. Lett., {\bf 72}, 486 (2005); M. Wyart, L.E. Silbert,
S.R. Nagel, T.A. Witten, {\it Phys. Rev. E}{\bf 72} 051306 (2005); C. Brito, M. Wyart, Europhys. Lett. {\bf 76} 149 (2006).
\bibitem{VanHecke} W.G. Ellenbroek, E. Somfai, M. van Hecke, W. van Saarloos,  {\it Phys. Rev. Lett.} {\bf 97}, 258001 (2006).
\bibitem{D'Anna} G. D'Anna, G. Gremaud, {\it Nature} {\bf 413}, 407  (2001).
\bibitem{Pouliquen} O. Pouliquen, M. Belzons, and M. Nicolas, {\it Phys. Rev. Lett.} {\bf 91} 014301 (2003).
\bibitem{Kabla} A. Kabla, G. Debr{\'e}geas, {\it Phys. Rev. Lett.} {\bf 85} 3632 (2000).
\bibitem{Marty} G. Marty, O. Dauchot {\it Phys. Rev. Lett.} {\bf 94}, 015701 (2005).
\bibitem{Reis}  P.M. Reis, R.A. Ingale, M.D. Shattuck,  {\it Phys. Rev. Lett.} {\bf 98}, 188301 (2007).
\bibitem{AbateDurian} A. R. Abate, D. J. Durian {\it Phys. Rev. E} {\bf 74}, 031308 (2006).
\bibitem{weeks} E. Weeks, J.C. Crocker, A.C. Levitt, A. Schofield, and D.A. Weitz, {\it Science} {\bf 287}, 627 (2000).
\bibitem{Dauchot} O. Dauchot, G. Marty, G. Biroli, {\it Phys. Rev. Lett.} {\bf 95} 265701 (2005).
\bibitem{ScienceBerthieretal} L. Berthier et al., {\it Science} {\bf 310}, 1797 (2005); L. Berthier et al., {\it J. Chem. Phys.} {\bf 126}, 184503 (2007).
\bibitem{Abate}  A. S. Keys, A. R. Abate, S. C. Glotzer, D. J. Durian, {\it Nature Physics} {\bf 3}, 260 (2007).
\bibitem{luca} L. Cipelletti and L. Ramos, {\it J. Phys.: Condens. Matter} {\bf 17}, R253 (2005). 
\bibitem{jorge} F. Krzakala, J. Kurchan, Phys. Rev. E 76, 021122 (2007), arXiv:0709.1023 and private communication.
\bibitem{semerjian} G. Semerjian, J. Stat. Phys. 130, 251 (2008).
\bibitem{Swinney} D. I. Goldman and H. L. Swinney, Phys. Rev. Lett. {\bf 96} 145702 (2006). 
\bibitem{Behringer}  T. S. Majmudar, M. Sperl, S. Luding, R. P. Behringer , {\it Phys. Rev. Lett.} {\bf 98} 058001 (2007).
\bibitem{Nagel} E. R. Nowak, J. B. Knight, E. Ben-Naim, H. M. Jaeger, and S. R. Nagel, {\it Phys. Rev. E} {\bf 57}, 1971 (1998).
\bibitem{FranzParisiBerthierHarrowell} C. Donati, S. Franz, G.
Parisi, and S.C. Glotzer, J. Non-Cryst. Solids  {\bf 307}, 215 (2002); S. Whitelam, L. Berthier, J.~P. Garrahan, 
{\it Phys. Rev. Lett.} {\bf 92}, 185705 (2004). 
\bibitem{leshouches} J.P. Bouchaud, Proceedings of the 2002 Les Houches Summer School on Slow Relaxations and
Nonequilibrium Dynamics in Condensed Matter; J.H. Snoeijer, T.J.H. Vlugt, M. van Hecke, and W. van Saarloos,
{\it Phys. Rev. Lett.} {\bf 92}, 054302 (2004).
\bibitem{garrahan} D. Chandler, J.P. Garrahan, R L. Jack, L. Maibaum, A.C. Pan, Phys. Rev. E 74, 051501 (2006)
\bibitem{EdwardsBertin} S. F. Edwards and R. B. S. Oakeshott, {\it Physica A}, {\bf 157} 1080, (1989); E. Bertin, K. Martens, O. Dauchot, and M. Droz, {\it Phys. Rev. E} {\bf 75}, 031120 (2007).
\bibitem{EPLBound} F. Lechenault, O. Dauchot, G. Biroli, J.-P. Bouchaud, {\it cond-mat/0712.2036}.
\bibitem{LiuGoree} B. Liu, J. Goree, {\it Phys. Rev. E} {\bf 75}, 016405 (2007).
\bibitem{Glotzer} C. Donati, S.C. Glotzer., P.H. Poole, W. Kob, S.J. Plimpton, {\it Phys. Rev. E} {\bf 60}, 3107 (1999).
\bibitem{zamponi} G. Parisi and F. Zamponi, arXiv:0802.2180.
\bibitem{Grigera} T. Grigera, A. Cavagna, I. Giardina and G. Parisi, {\it Phys. Rev. Lett.} {\bf 88} 055502 (2002); T. Grigera et al. {\it Nature} {\bf 422} 289 (2003).
\bibitem{BiroliBouchaud} T. Grigera, A. Cavagna, I. Giardina and G. Parisi, {\it Phys. Rev. Lett.} {\bf 97}, 195701 (2006).
\bibitem{SellittoBiroliToninelli} M. Sellitto, G. Biroli, and C. Toninelli,  {\it Europhys. Lett.} {\bf 69} 496 (2005).
\bibitem{Schwarz} J. Schwarz, A. J. Liu, L. Q. Chayes, {\it Europhys. Lett.} {\bf 73}, 570 (2006).
\bibitem{pitard} J.-P. Bouchaud, E. Pitard, {\it Eur. Phys. J. E} {\bf 6} 231 (2001).

\end{thebibliography}
\end{document}